\documentclass[5p,two column,sort&compress]{elsarticle}

\usepackage{graphicx,placeins,amsmath,amssymb,subfigure}

\def\fm{\mathrm{fm}}
\def\mev{\mathrm{MeV}}
\def\gev{\mathrm{GeV}}

\def\msbar{\overline{\mathrm{MS}}}
\def\lsim{\raise0.3ex\hbox{$<$\kern-0.75em\raise-1.1ex\hbox{$\sim$}}}
\def\gsim{\raise0.3ex\hbox{$>$\kern-0.75em\raise-1.1ex\hbox{$\sim$}}}

\newcommand{\GeV}{\,\mathrm{GeV}}

\newcommand{\temp}[1]{#1}
\begin{document}


\title{Precision computation of the kaon bag parameter}

\author[buw,fzj]{S.~D\"urr}
\author[buw,fzj,itp]{Z.~Fodor}
\author[buw]{C.~Hoelbling}
\author[buw,itp]{S.D.~Katz}
\author[buw,fzj]{S.~Krieg}
\author[buw]{T.~Kurth}
\author[cpt]{L.~Lellouch\fnref{cptaffil}}
\author[buw,fzj]{T.~Lippert}
\author[buw]{C.~McNeile}
\author[cpt]{A.~Portelli\fnref{cptaffil}}
\author[buw]{K.K.~Szab\'o}

\address[buw]{Bergische Universit\"at Wuppertal, Gaussstr.\,20, D-42119 Wuppertal, Germany.}
\address[fzj]{J\"ulich Supercomputing Centre, Forschungszentrum J\"ulich, D-52425 J\"ulich, Germany.}
\address[itp]{Institute for Theoretical Physics, E\"otv\"os University, H-1117 Budapest, Hungary.}
\address[cpt]{Centre de Physique Th\'eorique, Case 907, Campus de Luminy, F-13288 Marseille, France.}
\fntext[cptaffil]{CPT is research unit UMR 6207 of the CNRS and of the universities Aix-Marseille I, Aix-Marseille II and Sud Toulon-Var, and is affiliated with the FRUMAM.}

\date{\today}

\begin{abstract}

\temp{Indirect CP violation in $K\to\pi\pi$ decays plays a central role in constraining the flavor structure of the Standard Model (SM) and in the search for new physics. For many years the leading uncertainty in the SM prediction of this phenomenon was the one associated with the nonperturbative strong interaction dynamics in this process. Here we present a fully controlled lattice QCD calculation of these effects, which are described by the neutral kaon mixing parameter $B_K$. We use a two step HEX smeared clover-improved Wilson action, with four lattice spacings from $a{\approx} 0.054\,\fm$ to $a{\approx} 0.093\,\fm$ and pion masses at and even below the physical value. Nonperturbative renormalization is performed in the
  RI-MOM scheme, where we find that operator mixing induced by chiral
  symmetry breaking is very small. Using fully nonperturbative
  continuum running, we obtain our main result
  $B_K^\mathrm{RI}(3.5\,\mathrm{GeV}){=}
  0.531(6)_{\mathrm{stat}}(2)_{\mathrm{sys}}$. A perturbative 2-loop
  conversion yields
  $B^{\msbar\mathrm{-NDR}}_K(2\,\gev)=0.564(6)_{\mathrm{stat}}(3)_{\mathrm{sys}}(6)_{\mathrm{PT}}$}
and $\hat{B}_K=0.773(8)_{\mathrm{stat}}(3)_{\mathrm{sys}}(8)_{\mathrm{PT}}$, which is in good agreement with current results from fits to experimental data.
\end{abstract}

\maketitle

\section{Introduction} 
Neutral kaon mixing is responsible for indirect CP-violation in $K{\rightarrow}\pi\pi$ decays. This violation is quantified by the parameter $\epsilon$, which is related to
quark flavor mixing parameters and the ratio of hadronic matrix elements
\begin{equation}\label{bk_definition}
B_K=\frac{\langle \bar{K}^0| O^{\Delta S=2}|K^0\rangle}{\frac{8}{3}\langle \bar{K}^0|A_\mu|0\rangle\langle 0|A^\mu|K^0\rangle},
\end{equation}
where $O^{\Delta S=2}=[\bar{s}\gamma_\mu(1-\gamma_5)d][\bar{s}\gamma^\mu(1-\gamma_5)d]$ (cf. \cite{Colangelo:2010et} for details). The computation of (\ref{bk_definition}) has some advantages over direct computations of the $O^{\Delta S=2}$ matrix element, such as the partial cancellation of statistical and systematic uncertainties.\\
Note that a precise determination of $B_K$ together with experimental measurements of $\epsilon$ yields important constraints on the unitary triangle parameters $(\bar{\rho},\bar{\eta})$.\\

\section{Lattice Details} 
We use the $N_f{=}2{+}1$ and $N_f{=}3$, 2 HEX  \cite{Capitani:2006ni} smeared, tree-level clover improved Wilson ensembles generated for determining light quark masses \cite{Durr:2010vn,Durr:2010aw}. 
Out of the five available lattice spacings, we use the four finest covering a range of $0.054\,\fm\,\lsim\, a\, \lsim\, 0.093\,\fm$ (\temp{the low momentum cutoff at the largest lattice spacing $a{\approx}0.116\,\mathrm{fm}$ does not allow for a reliable extraction of the mixing coefficients}). For the $N_f{=}2{+}1$ ensembles, the pion masses straddle the physical value.
Our lattices have sizes as large as $L{\sim} 6\,\fm$ and finite volume corrections to $M_\pi$ \cite{Colangelo:2005gd} are below 0.5\% \cite{Durr:2010aw}. \temp{We also computed the finite volume corrections to $B_K$ using the results from \cite{Becirevic:2003wk}. We found that these effects are even below the $0.3\%$ level and thus fully under control.} The $N_f{=}3$ configurations are used to compute the required renormalization constants nonperturbatively using the RI-MOM method \cite{Martinelli:1994ty,Donini:1999sf}.\\

\section{Nonperturbative renormalization}
\begin{figure}
\centering
\includegraphics[scale=0.36]{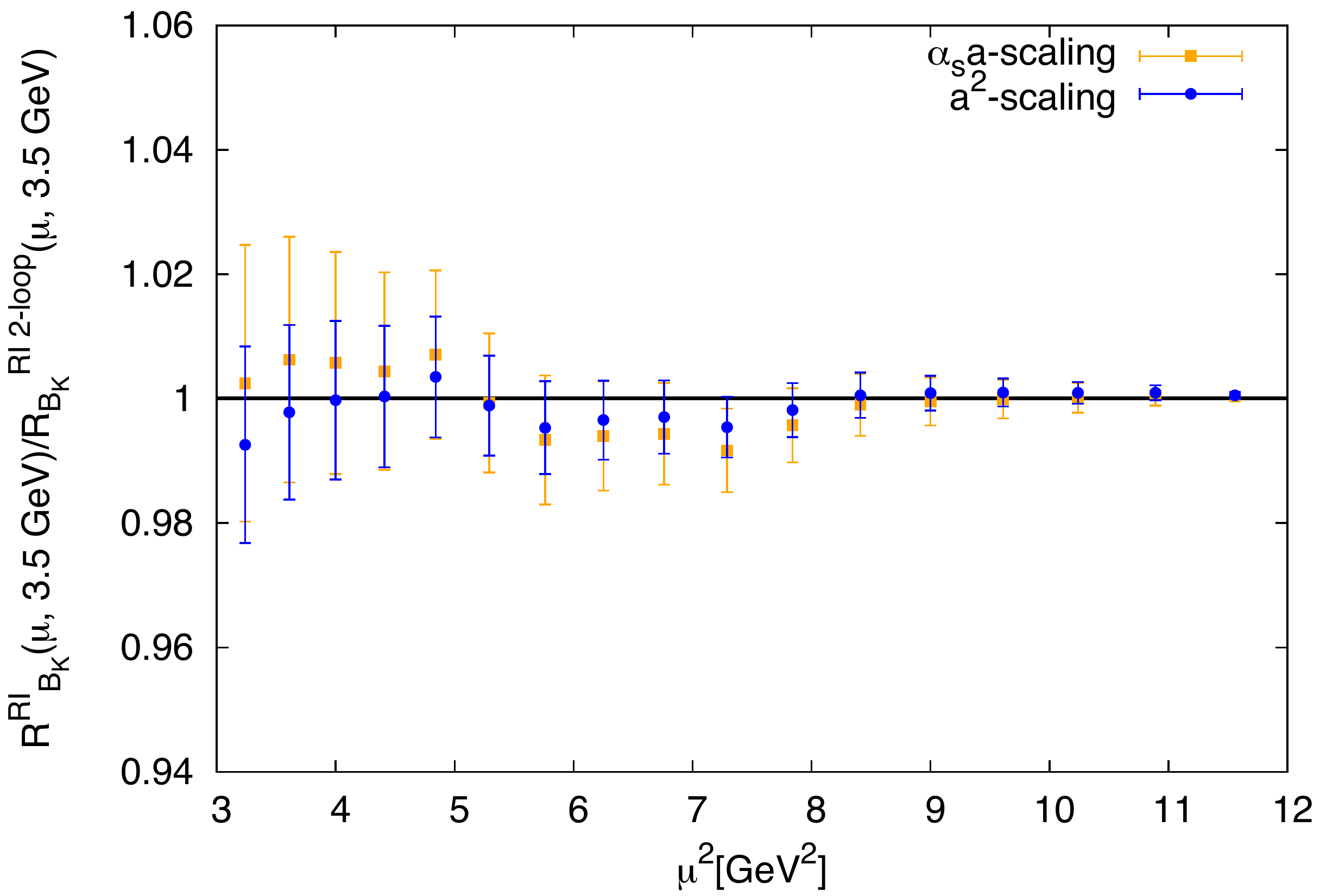}
\caption{\label{contlimit_pt_matched}Nonperturbative running obtained from the continuum extrapolation of (\ref{zbk_ratio}) assuming $\mathcal{O}(\alpha_s a)$ (orange) or $\mathcal{O}(a^2)$ (blue) scaling (for details see \cite{Durr:2010aw}), divided by the same running computed at NLO. We observe agreement of the runnings between $\mu=1.8\,\gev$ and $3.5\,\gev$ within a statistical error which grows to 2\% at the lower end of the range.}
\end{figure}

\begin{figure*}
\centering
\includegraphics[scale=0.38]{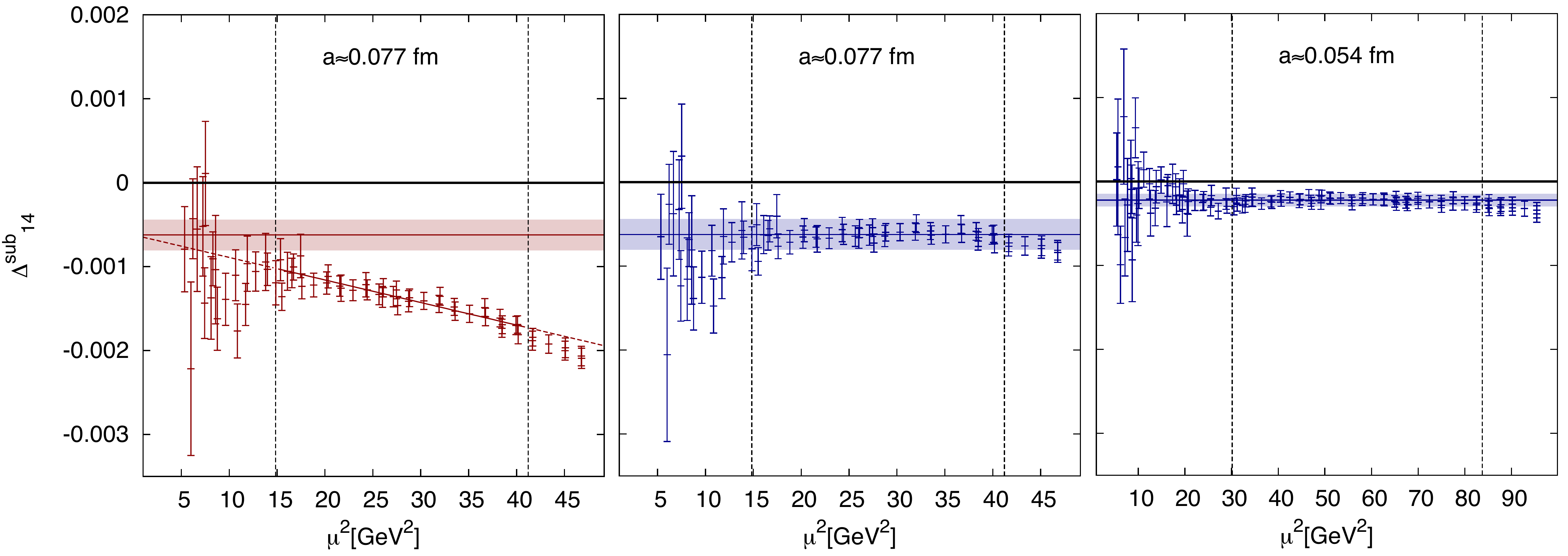}
\caption{\label{mixing_terms} Pole subtracted mixing term $\Delta^{\mathrm{sub}}_{14}$ as defined in (\ref{mix_sub_def}) and extrapolated to the chiral limit at $a{\approx} 0.077\,\fm$ lattice spacing (left). The next panel (middle) shows the same data with the $\mathcal{O}((ap)^2)$ discretization error removed. The same procedure was applied to $\Delta_{14}^{\mathrm{sub}}$ at $a{\approx} 0.054\,\fm$ (right). The dashed vertical bars indicate the corresponding fitting regions and the horizontal line corresponds to the extracted $\Delta_{14}$ along with its statistical $1\sigma$ error band. Note the long plateaus in which the data agree with the fit. The extracted mixing coefficient tends to zero as $a$ is reduced.}
\end{figure*}

Due to explicit chiral symmetry breaking of the Wilson action, the parity even part of operator $O^{\Delta
  S=2}=(V-A)(V-A)= O_1$ mixes with the other dimension six operators 
$O_{2}=VV-AA$, $O_{3/4}=SS\mp PP$ and  $O_5=TT$
where $V,A,S,P,T$ denote vector, axial-vector, scalar, pseudoscalar and tensor $\Delta S=1$ bilinears respectively. We denote their bare matrix elements $\langle \bar{K}^0|O_i|K^0\rangle $ by $Q_i$. The renormalization pattern is then given by \cite{Donini:1999sf}
\begin{equation}
Q^{\mathrm{ren}}_i=\tilde{Z}_{ik}Q_k=Z_{ij}(\delta_{jk}+\Delta_{jk})Q_k.
\end{equation}
Hence, the renormalization matrix $\tilde{Z}_{ij}$ is decomposed into $Z_{ij}$, which, analogously to the continuum renormalization, only mixes $O_{2/3}$ and $O_{4/5}$ respectively, and a correction $\Delta_{jk}$, quantifying the mixing of different $O_k$ due to chiral symmetry breaking. Since $O_1$ does not mix in the continuum, the only relevant terms in the above expression are $Z_{11}$ and $\Delta_{1k},\mbox{ for }k=2,\ldots,5$. Because of (\ref{bk_definition}), the multiplicative renormalization factor for $B_K$ is given by $Z_{B_K}= Z_{11}/Z_A^2$.\\
We use the nonperturbative running method \cite{Constantinou:2010gr,Arthur:2010ht,Durr:2010aw} to circumvent the window-problem of RI-MOM \cite{Martinelli:1994ty,Giusti:2000jr} and allow matching to other schemes such as $\msbar$ and RGI with small perturbative uncertainties. This means that, after an essentially flat extrapolation of $Z_{B_K}$ to vanishing quark mass for each lattice spacing, we compute the ratio
\begin{equation}\label{zbk_ratio}
R^\mathrm{RI}_{B_K,\beta}(\mu,3.5\,\gev)=\frac{Z^\mathrm{RI}_{B_K,\beta}(3.5\,\gev)}{Z^\mathrm{RI}_{B_K,\beta}(\mu)}
\end{equation}
on the three finest lattices at different $\mu$ between $1.8\,\gev$ and $3.5\,\gev$. This ratio is then extrapolated to the continuum limit yielding the nonperturbative running factor from $\mu$ to $3.5\,\gev$, $R^\mathrm{RI}_{B_K}(\mu,3.5\,\gev)$. Fig. \ref{contlimit_pt_matched} shows that our continuum extrapolated results for the running agree with NLO perturbation theory \cite{Ciuchini:1997bw,Buras:2000if} in the $\mu$-range considered. 
Thus we set $Z^\mathrm{RI}_{B_K,\beta}(3.5\,\gev)= R^\mathrm{RI}_{B_K}(\mu,3.5\,\gev)\cdot Z^\mathrm{RI}_{B_K,\beta}(\mu)$, where $\mu$ is chosen such that $Z^\mathrm{RI}_{B_K,\beta}(\mu)$ can be safely extracted for all four lattice spacings. As an additional improvement, we subtract a contact term from the propagators as described in \cite{Becirevic:1999kb,Capitani:2000xi,Maillart:2008pv}.\\
The mixing coefficients $\Delta_{1k}$ are obtained as described in \cite{Donini:1999sf}, where the additional subtraction \cite{Giusti:2000jr}
\begin{equation}\label{mix_sub_def}
\Delta^{\mathrm{sub}}_{1k}(a,m_1,m_2)=\frac{m_1\Delta_{1k}(a,m_1)-m_2\Delta_{1k}(a,m_2)}{m_1-m_2}
\end{equation}
is applied. Here, $\Delta_{1k}(a,m_i)$ is the mixing coefficient obtained at quark mass $m_i$. This procedure removes $\mathcal{O}(p^{-2})$ contributions coming from virtual pion exchanges. The dominant corrections are then $\mathcal{O}((ap)^2)$ discretization errors and an $\mathcal{O}(p^{-4})$ term attributed to double pion exchanges. We use different fit windows as well as fit functions, which include either an $(ap)^2$ term or an additional $p^{-4}$ term to estimate systematic effects coming from this ambiguity. However, all these effects turn out to be very small, as both fit functions give compatible results. We also remove the small remaining quark mass dependence in the same fit. Fig. \ref{mixing_terms} shows the mixing term $\Delta^{\mathrm{sub}}_{14}$ prior to and after removing the discretization effects at $a{\approx}0.077\,\mathrm{fm}$ lattice spacing. The same data at $a{\approx}0.054\,\mathrm{fm}$ are also shown for comparison. We observe that all mixing coefficients are small and tend to zero as $a$ is decreased.\\\\

\section{Matrix Elements}
\begin{figure}[h]
\centering
\includegraphics[scale=0.7]{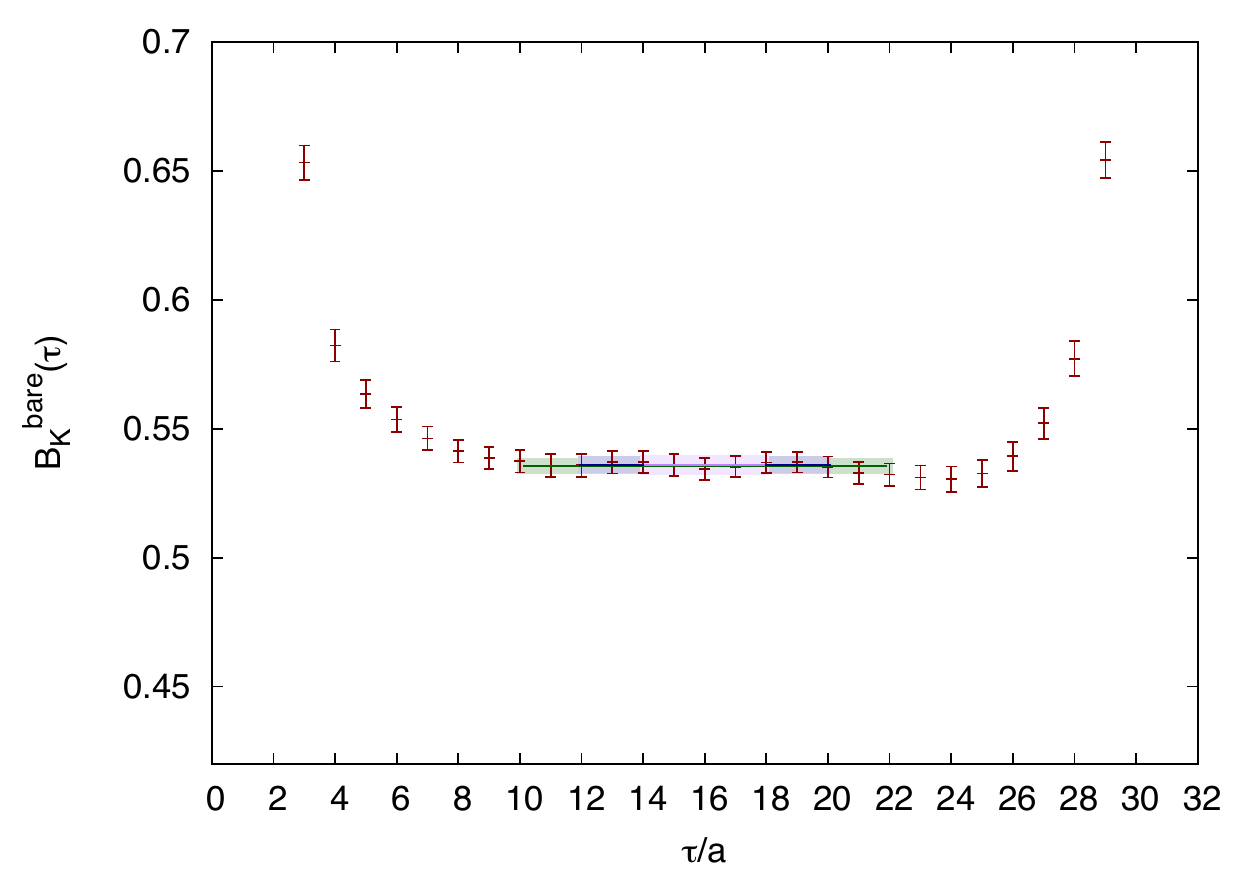}
\caption{\label{bk_plateaus}Plateau for $B_K^\mathrm{bare}$ (see expression (\ref{bk_definition})) at \temp{$a{\approx}0.065\,\mathrm{fm}$, with $M_\pi{\sim}245\,\mev$ and time extent $T/a{=}64$, symmetrized with respect to $t{=}T/2$}. Different solid lines represent constant fits for different fitting ranges and the shaded bands the corresponding statistical error determined on 2000 bootstrap samples. All fits to the different ranges agree very well within these errors.}
\end{figure}

\begin{figure}
\centering
\includegraphics[scale=0.725]{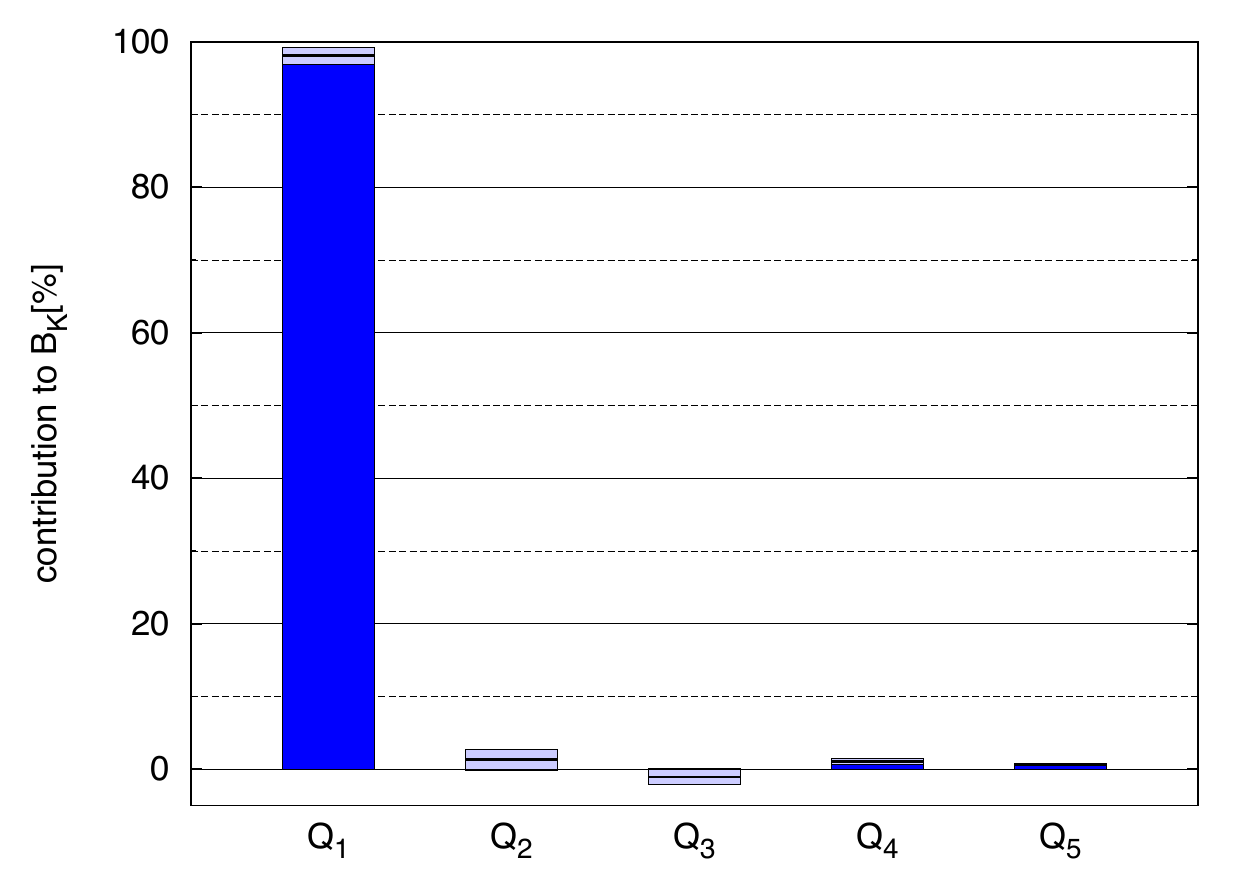}
\caption{\label{bk_budget} Total contribution of individual $Q_1,\Delta_{1i}Q_i$ to $B_K$ in \% for $a{\approx} 0.077\,\mathrm{fm}$ and $M_\pi{\sim} 120\,\mathrm{MeV}$ with their errors. The contributions from operators $Q_2$ to $Q_5$ are small or even compatible with zero. }
\end{figure}

To obtain the matrix elements relevant for the computation of $B_K$, we use color-random $\mathrm{U(1)}$ wall sources
at $t=0$ and $t=T/2$ \cite{Bitar:1988bb} and vary the time slice
$\tau$ of the operator insertion between $1$ and $T-1$. The relevant
operator insertions are $O_{1\ldots 5}$. The matrix elements $Q_i$ are determined by performing constant fits of the time-symmetrized plateaus in $\tau$ as shown in Fig. \ref{bk_plateaus}. We use three different fitting ranges in order to estimate systematic effects due to excited states.
Combining the results from these fits and the $\Delta_{1k}$ determined before, we can decompose $B_K$ into the contributions from the individual $Q_i$. 
The chiral symmetry breaking contributions of operators $O_{2,\ldots,5}$ are $\mathcal{O}(\alpha_s)$ suppressed relative to that of the leading operator $O_1$. However, the corresponding matrix elements $Q_{2,\ldots,5}$ are chirally enhanced and grow relative to $Q_1$ as the $\mathrm{SU(3)}$ chiral limit is approached. Thus it is important to control the subtraction of these chiral symmetry breaking contaminations. The good chiral properties of our fermion action mitigate this problem since the contribution of $Q_1$ largely dominates. As shown in Fig. \ref{bk_budget}, it is $98.1(1.2)\%$ for $a{\approx} 0.077\,\mathrm{fm}$, $M_\pi{\sim} 120\,\mathrm{MeV}$ and $m_s$ very close to its physical value.

\section{Extraction of Physical $B_K$} 
We perform a combined chiral and continuum fit to extract the renormalized $B_K$ at the physical mass point and in the continuum limit. Since we simulate at or below the physical light quark masses, we can perform a safe interpolation in the quark masses instead of relying on extrapolation formulas. For this combined fit we choose the following functional form
\begin{equation}
B^{\mathrm{RI}}_K(3.5\,\gev,x,y,a)=B^{\mathrm{RI}}_K(3.5\,\gev)\cdot f(x,y)+d(a),
\end{equation}
where $f(x,y)$ with $x{=}M_\pi^2$ and $y{=}2M_K^2-M_\pi^2$ describes the quark mass dependence. The generic form of $f$ is
\begin{align}
f(x,y)&=\left(1+a_{10}x+a_{20}x^2+a_{01}y\right.\nonumber\\&\left.+a_{11}xy-\frac{a_\chi x}{32\pi^2 f_0^2}\log\left(\frac{x}{\mu^2}\right)\right).
\end{align}
We use in total five different fit forms: three Taylor fits with different powers in $x$ and $y$, i.e. with $a_\chi{=}0$ and $a_{10},a_{01}$ left free as well as either $a_{20}$ or $a_{11}$ free or set to zero.
Additionally, we apply two SU(2) $\chi$PT fits
(cf. \cite{Constantinou:2010qv,Aoki:2010pe}) with $a_\chi{=}1$ and
$a_{10}=a_{20}=0$ as well as $f_0,\mu,a_{01}$ free and $a_{11}$ either left free or kept fixed to zero. All fits have good fit quality and show full
agreement with the expected
SU(2) chiral behavior for the case
$a_\chi{=}1$.
Since the ratio (\ref{bk_definition}) is tree-level $O(a)$ improved,
$d(a)$ is chosen proportional to either $\alpha_s a$ or $a^2$, as
discussed in \cite{Durr:2010aw}. We do not include terms whose coefficients are compatible
with zero in our final fits. In addition, we use the expressions given in
\cite{Becirevic:2003wk} to correct the data for the remaining small
finite volume effects. We further apply two different pion mass cuts of $380\,\mev$ and $340\,\mev$ (cf. \cite{Durr:2010aw}).
\begin{figure}[ht]
\centering
\subfigure[]{
\includegraphics[scale=0.35]{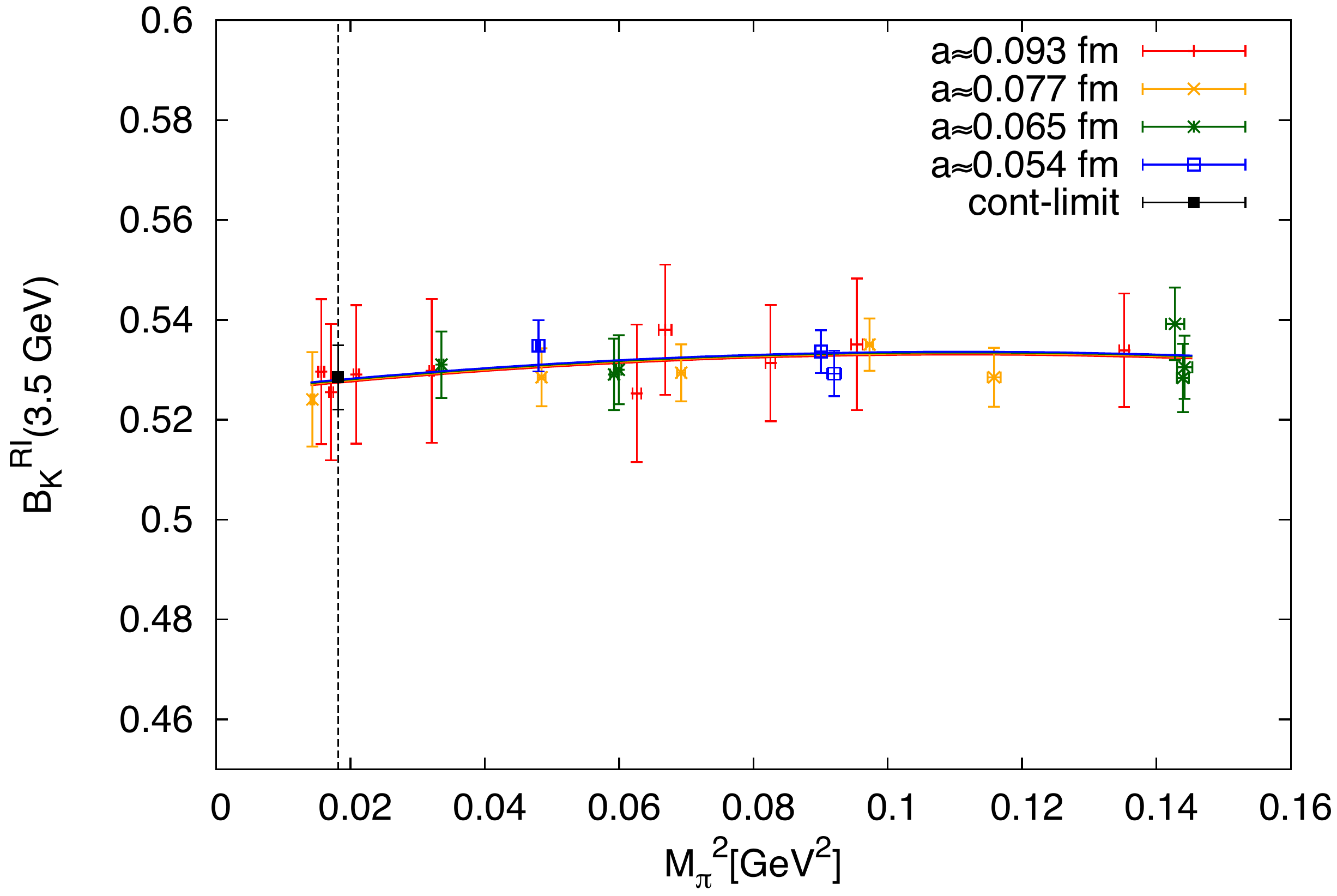}
}
\subfigure[]{
\includegraphics[scale=0.35]{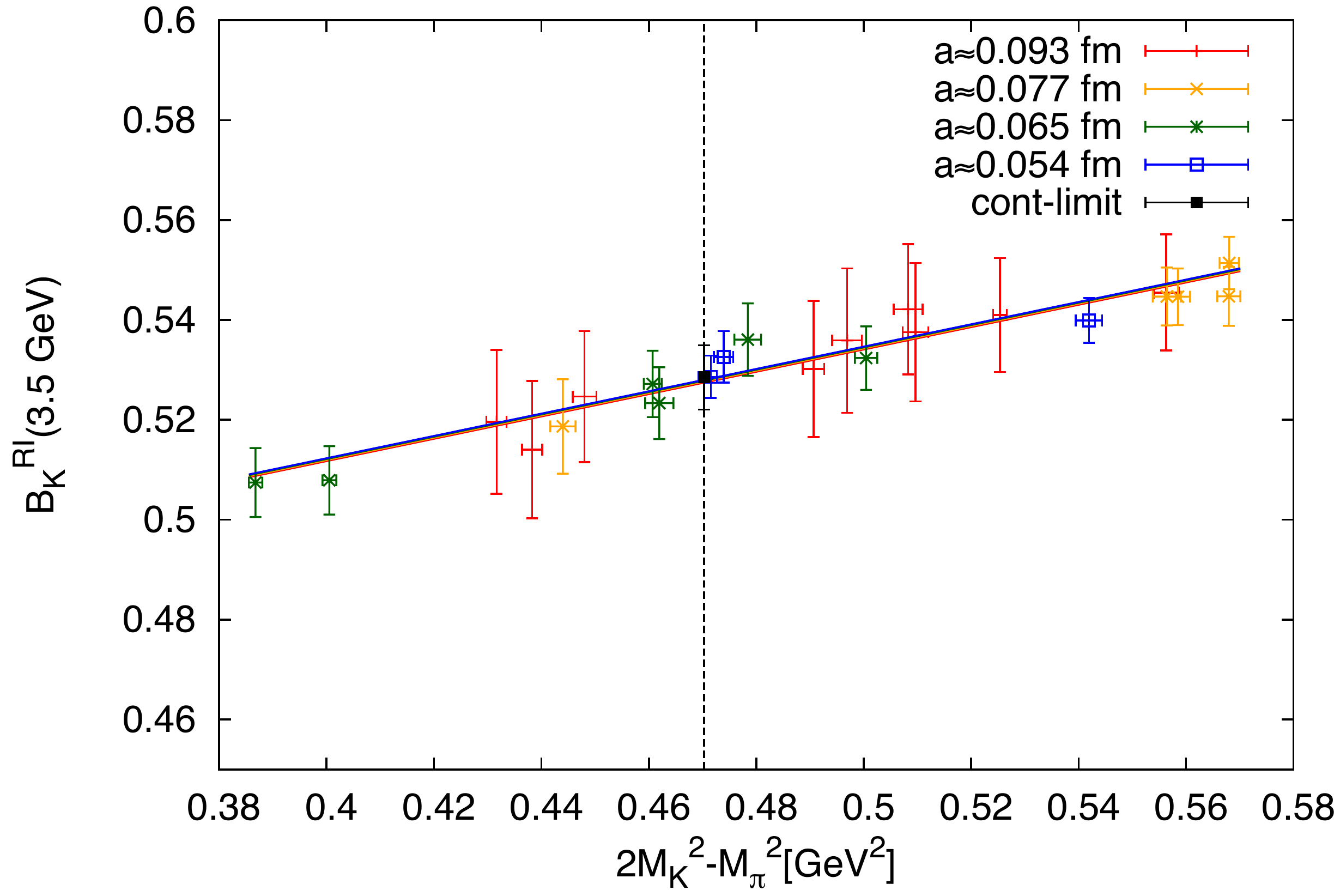}
}
\caption{\label{bk_combfit}\temp{Combined continuum extrapolation and interpolation to physical quark mass for a typical Taylor fit out of our 5760 fits. The filled square represents our result for $B_K^{\mathrm{RI}}(3.5\,\gev)$, the dashed vertical line the corresponding physical mass. As can be seen from panel (a), the interpolation in the light quark mass is mild. The slope of the interpolation in $m_s$ is somewhat steeper (b), but both interpolations are fully under control.}}
\end{figure}
\begin{figure}[ht]
\centering
\includegraphics[scale=0.35]{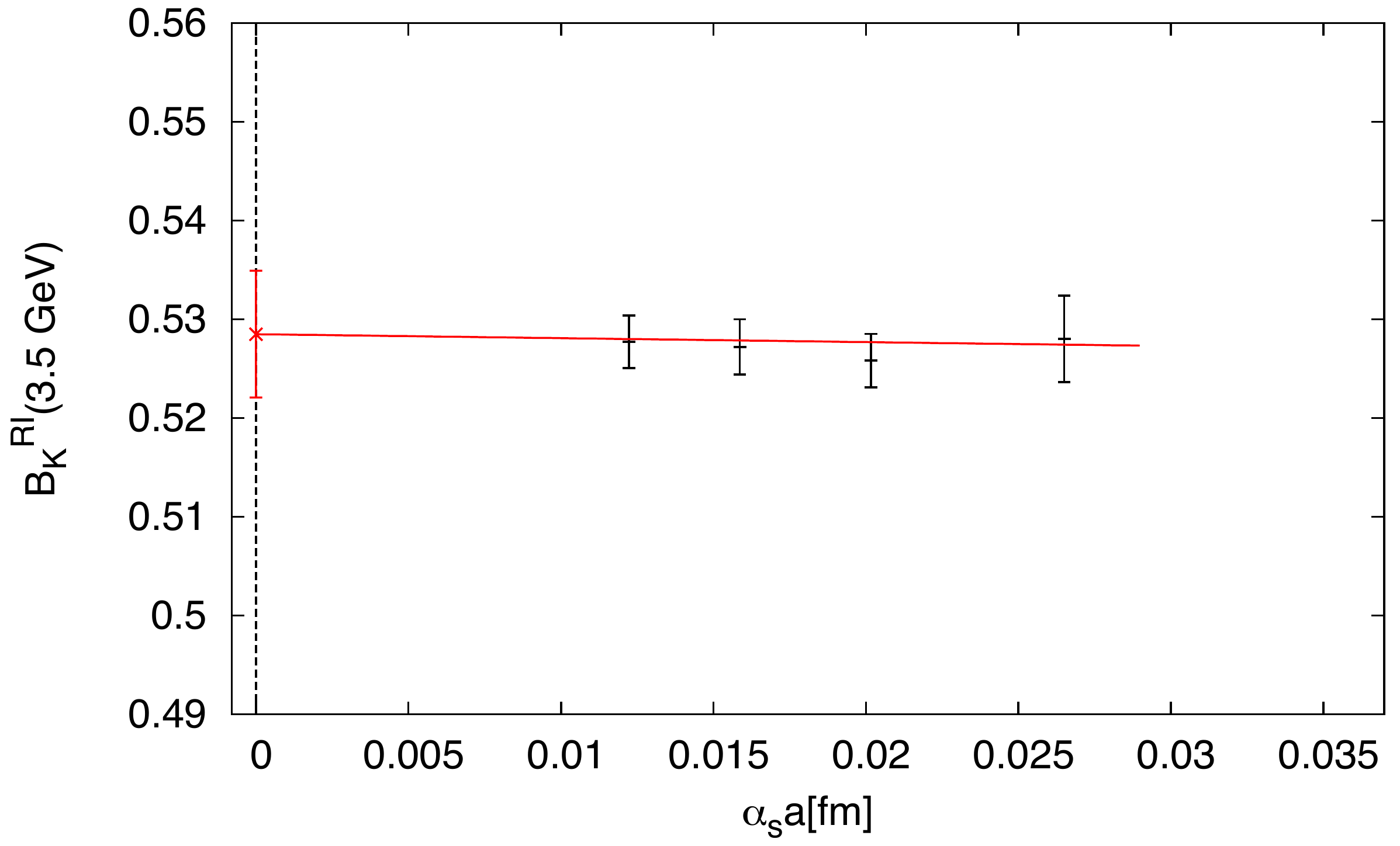}
\caption{\label{bk_contlimit}\temp{Continuum extrapolation of $B_K^{\mathrm{RI}}(a,3.5\,\gev)$, as obtained from the fit in Fig. \ref{bk_combfit}}.}
\end{figure}
\begin{figure}[ht]
\centering
\includegraphics[scale=0.8]{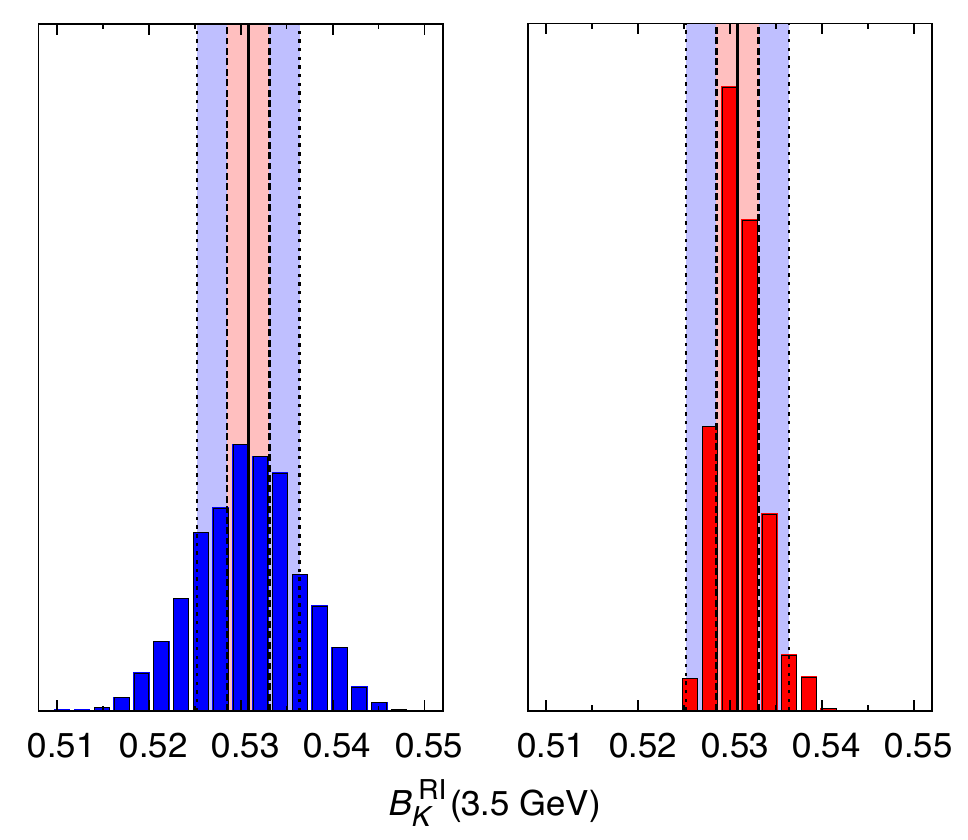}
\caption{\label{bk_hist}Statistical distribution (left) and the distribution attributed to systematic uncertainties (right) for $B_K^{\mathrm{RI}}(3.5\,\gev)$. The solid vertical line denotes our final value and the outer and inner bands our statistical and systematic errors. This figure emphasizes the fact that our overall error is completely dominated by the statistical uncertainties.}
\end{figure}
Together with two different fit ranges for pion and kaon mass extractions, different fit ranges and
functions for obtaining the mixing terms and three different scales
for extracting the renormalized matrix elements, we end up with 5760 different analyses. All of our fit results are very precise and
compatible with each other. 
A sample (Taylor) fit with $\alpha_s a$
scale dependence in Fig. \ref{bk_combfit} shows an essentially flat
chiral behaviour and extremely small discretization effects.
This is also evident from the continuum extrapolation shown in Fig. \ref{bk_contlimit}.
Using the method from
\cite{Durr:2008zz,Durr:2010aw} for a controlled determination of all systematics as well as the statistical error, our full nonperturbative main result reads
\begin{equation}
B_K^{\mathrm{RI}}(3.5\,\mathrm{GeV})=0.5308(56)_{\mathrm{stat}}(23)_{\mathrm{sys}},
\label{eq:bkrires}\end{equation}
where the individual contributions to the systematic error originate from the subtraction of the mixing terms ($0.0021$), excited state uncertainties ($0.0007$), extrapolating to the continuum and interpolating to physical quark masses (both $0.0006$), as well as the extraction of the renormalization constant ($0.0002$). Fig. \ref{bk_hist} shows the statistical and systematic error distributions of our values for $B_K^{\mathrm{RI}}(3.5\,\gev)$. Both distributions are fairly symmetric and clearly show that our final result is dominated by statistical uncertainties.

For the reader's convenience, we convert our main result of
(\ref{eq:bkrires}) into the $\msbar$-NDR scheme and into the RGI value
$\hat{B}_K$. We do so by using the NLO anomalous dimensions of
\cite{Ciuchini:1997bw,Buras:2000if} and the beta function at the
highest available loop order \cite{vanRitbergen:1997va}. It is
notoriously difficult to reliably assess the truncation error of a
perturbative series, particularly in the 68\% probability sense of our
systematic error treatment. As the NLO contributions to the conversion
factors are $\lesssim 2\%$ and NNLO contributions are typically much
smaller \cite{Buchalla:1995vs}, we add a rather conservative $1\%$
truncation error, which is larger than a variety of perturbative
estimates that we have tried.  Because this truncation error does not
fall into our fully controlled systematic error framework, we list it
separately and do not combine it with other systematics. We thus
obtain
\begin{align}
B^{\msbar\mathrm{-NDR}}_K(2\,\GeV)&=0.5644(59)_{\mathrm{stat}}(25)_{\mathrm{sys}}(56)_{\mathrm{PT}}\\
\label{eq:rgi}
\hat{B}_K&=0.7727(81)_{\mathrm{stat}}(34)_{\mathrm{sys}}(77)_{\mathrm{PT}}.
\end{align}

\begin{figure}[h!]
\centering
\includegraphics[scale=0.7]{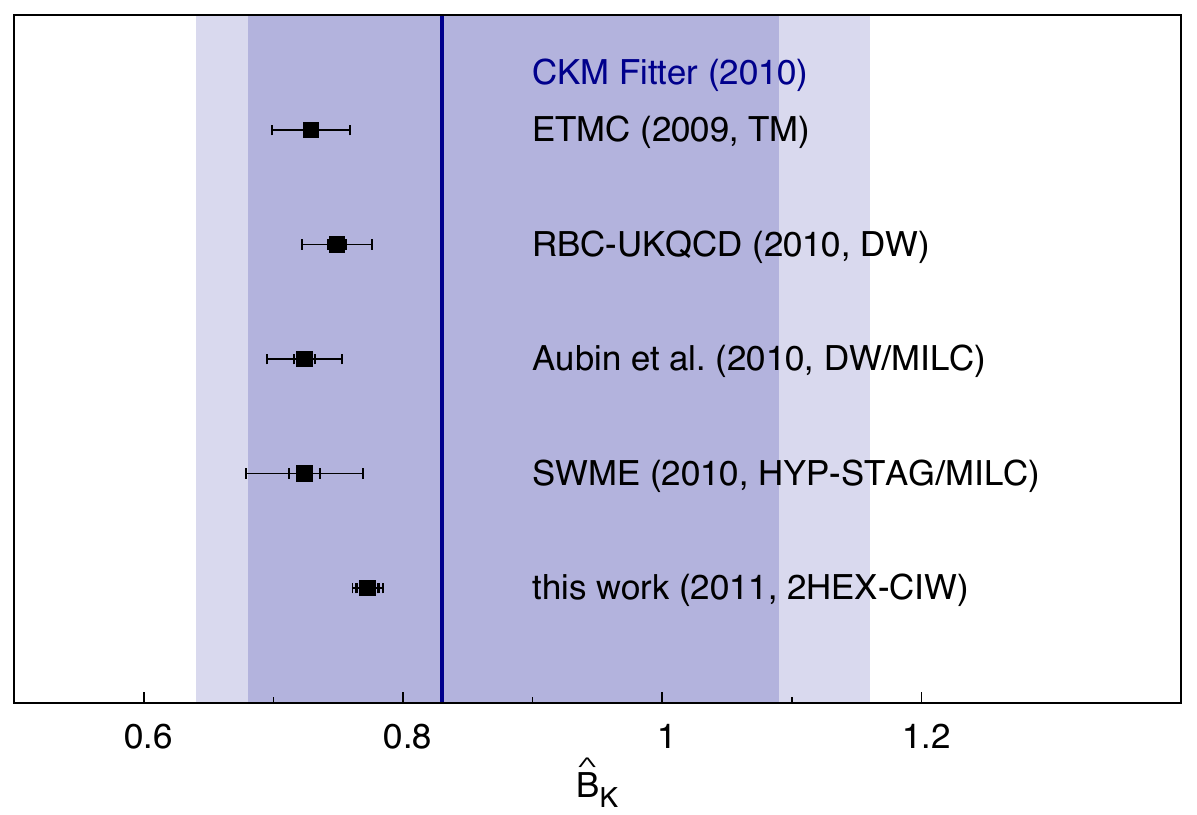}
\caption{\label{bk_compare}Comparison of our result (\ref{eq:rgi}) with the value for $\hat{B}_K$ obtained by CKMfitter (ICHEP 10 update to \cite{Charles:2004jd}, vertical line). The dark and light bands correspond to CKMfitter's $1\sigma$ and  $2\sigma$ confidence intervals, respectively. The results from different $N_f{=}2$ (\cite{Constantinou:2010qv}, 1st) and $N_f{=}2{+}1$ (\cite{Aubin:2009jh,Aoki:2010pe,Bae:2010ki}, 2nd to 4th) lattice computations are also shown.}
\end{figure}
In Fig.~\ref{bk_compare} we compare our result to Standard Model
expectations and other recent lattice results. 
Our result is in good agreement with indirect $B_K$ determinations from
global Standard Model fits of flavor mixing data obtained by CKMfitter (ICHEP 10 updates to \cite{Charles:2004jd}). It is consistent with expectations obtained by UTfit \cite{Bona:2006ah} by either including all decays ($\hat{B}_{K,\mathrm{all}}{=}0.94(17)$) or neglecting the semileptonic channels ($\hat{B}_{K,\mathrm{no-sl}}{=}0.88(13)$) in the fits. Therefore, we find no evidence for new fundamental contributions to indirect CP-violation in $K{\rightarrow}\pi\pi$ decays. This is in-line with the findings of \cite{Lunghi:2010gv}. Moreover, we hope that the high precision of our result will encourage our colleagues responsible for the determination of the other contributions to epsilon to work on reducing their uncertainties.\\

\textbf{Acknowledgments.} We thank J\'er\^ome Charles, Robert Harlander and Steve Sharpe for helpful discussions. Computations were performed using HPC resources from
FZ J\"ulich and from GENCI-[IDRIS/CCRT] (grant 52275) and
clusters at Wuppertal and CPT.  This work is supported in part by EU
grants I3HP, FP7/2007-2013/ERC n$^o$208740, MRTN-CT-2006-035482
(FLAVIAnet), DFG grant FO 502/2, SFB-TR 55, by CNRS grants GDR n$^0$
2921 and PICS n$^0$ 4707.

\bibliographystyle{elsarticle-num}
\bibliography{bk_plb}

\end{document}